\documentclass[sigconf,natbib=false]{acmart}

\AtBeginDocument{%
  }

\usepackage{tikz}
\usepackage{amsmath}
\usepackage{ulem}
\usepackage{amsmath,amsthm,amsfonts,amssymb,bm}
\usepackage[subrefformat=parens,labelformat=parens]{subfig}
\usepackage{graphicx}
\usepackage{pifont}
\usepackage{textcomp}
\usepackage{xcolor}
\usepackage{colortbl}
\usepackage{hhline}
\usepackage{algorithm}
\usepackage{algpseudocode}
\usepackage{hyperref}
\usepackage{array}
\usepackage{setspace}
\usepackage{colortbl}
\usepackage{booktabs}
\usepackage{multirow}
\usepackage{comment}
\usepackage{threeparttable}
\usepackage{caption}
\usepackage{tabularx}

\RequirePackage[
  datamodel=acmdatamodel,
  style=acmnumeric,
  maxbibnames=1,
  ]{biblatex}

\DeclareFieldFormat{doi}{}
\DeclareFieldFormat{eventtitle}{}

\renewcommand\footnotetextcopyrightpermission[1]{}
\DefineBibliographyStrings{english}{
  andothers = {et\addabbrvspace al.}
}

\definecolor{dkgreen}{rgb}{0,0.6,0}
\definecolor{mauve}{rgb}{0.58,0,0.82}
\newcommand{\commentOutputSwitch}{0} 
\ifnum\commentOutputSwitch=1
    \newcommand{\ml}[1]{{\color{red}\bf [Meng: #1]}}
    \newcommand{\gqy}[1]{{\color{blue}\bf [GQY: #1]}}
\else
    \newcommand{\ml}[1]{}
    \newcommand{\gqy}[1]{}
    \renewcommand{\sout}[1]{}
\fi
\newcommand{\red}[1]{\textcolor{red}{#1}}   % define text color red to be a simpler command
\newcommand{\blue}[1]{\textcolor{blue}{#1}}
\newcommand{\green}[1]{\textcolor{dkgreen}{#1}}

\newcommand{\method}{HG-PIPE}

\begin{document}

\fontsize{9pt}{10.7pt}\selectfont

\title{HG-PIPE: Vision Transformer Acceleration with Hybrid-Grained Pipeline}

\author{
  Qingyu Guo\textsuperscript{1}, Jiayong Wan\textsuperscript{3}, Songqiang Xu\textsuperscript{3}, Meng Li\textsuperscript{2,1,4*}, Yuan Wang\textsuperscript{1,4*}
  \vspace{0.1em}
}
\affiliation{
  \institution{\textsuperscript{1}School of Integrated Circuits, Peking University, China  \hspace{1.2em}  \textsuperscript{2}Institute for Artificial Intelligence, Peking University, China}\country{}
}
\affiliation{
  \institution{\textsuperscript{3}School of Software and Microelectronics, Peking University, China}\country{}
}
\affiliation{
  \institution{\textsuperscript{4}Beijing Advanced Innovation Center for Integrated Circuits, China}\country{}
}
\affiliation{
  % \institution{\textsuperscript{*}meng.li@pku.edu.cn, wangyuan@pku.edu.cn}\country{}
  \institution{{*}\textit{meng.li@pku.edu.cn}, {*}\textit{wangyuan@pku.edu.cn}}\country{}
}

%%
%% By default, the full list of authors will be used on the page
%% headers. Often, this list is too long and will overlap

%-------------------------------------------------------------------------------
\begin{abstract}
%-------------------------------------------------------------------------------
Vision Transformer (ViT) acceleration with field programmable gate array (FPGA)
is promising but challenging.
Existing FPGA-based ViT accelerators mainly rely on temporal architectures,
which process different operators by reusing the same hardware blocks and suffer
from extensive memory access overhead.
Pipelined architectures, either coarse-grained or fine-grained,
unroll the ViT computation spatially for memory access efficiency.
However, they usually suffer from significant hardware
resource constraints and pipeline bubbles induced by the global computation dependency of ViT.
In this paper, we introduce \method, a pipelined FPGA accelerator for high-throughput and low-latency ViT processing.
\method~features a hybrid-grained pipeline architecture to reduce on-chip buffer cost and 
couples the computation dataflow and parallelism design to eliminate the
pipeline bubbles.
\method~further introduces careful approximations to implement both linear and 
non-linear operators with abundant Lookup Tables (LUTs), thus alleviating resource constraints.
On a ZCU102 FPGA, \method~achieves 2.78$\times$ better throughput and 2.52$\times$ better resource efficiency than the prior-art accelerators, e.g., AutoViTAcc.
With a VCK190 FPGA, \method~realizes end-to-end ViT acceleration on a single device and achieves 7118 images/s, which is 2.81$\times$ faster than a V100 GPU. 

\end{abstract}


\maketitle

\pagestyle{fancy}
\fancyhead{} % 清空页眉

\section{Introduction}
\label{sec: intro}

Recent years have witnessed the wide adoption of Transformer models in the field of computer vision (CV)\cite{vit, swin, yolo, detr, unet, segmenter}.
% particularly image classification~\cite{vit, swin}, object detection~\cite{yolo, detr}, semantic segmentation~\cite{unet, segmenter}, etc.
While Vision Transformers (ViTs) achieve state-of-the-art (SOTA) performance compared to convolutional neural networks (CNNs),
they suffer from a drastic increase in parameters and computation, which calls for more efficient acceleration.

\begin{figure}[!tb]
  \centering
  \includegraphics[width=0.5\textwidth]{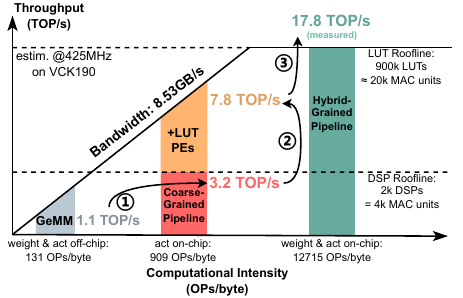}
  \caption{Roofline model for FPGA-based ViT acceleration.}
  % TODO: don't use caption, use \Description
  \Description{Roofline model for FPGA-based ViT acceleration.}
  \label{fig: roofline model}
\end{figure}

ViT acceleration based on field-programmable gate array (FPGA) has been actively
studied considering its efficiency and programmability \cite{FPGAPrinciples}.
FPGA-based ViT accelerators can be categorized into two architectural paradigms,
i.e., temporal architecture \cite{heat-vit, high_freq_sa, auto-vit-acc, VAQF, group_vector_sa} and pipelined architecture \cite{via, length_adaptive, TranCIM, ssr}.
Temporal architectures build general processing engines (PEs) for different
operators and reuse the PEs temporally \cite{high_freq_sa, group_vector_sa}.
Although temporal architectures simplify the design, they suffer from extensive
off-chip memory access\cite{defines} and utilization problems.
As shown in Figure \ref{fig: roofline model}, the temporal architecture (labeled "GeMM") is limited by bandwidth and can only achieve 1.1 TOP/s in our estimation.
Pipelined architectures, in contrast, instantiate distinct PEs specialized for
different operators and directly stream activations among PEs to reduce
off-chip memory access \cite{via, FixyFPGA, tomato, ssr}. Since they allow concurrent processing of multiple
layers, pipeline architectures hold the promise to enable efficient and low-latency ViT acceleration.

However, to realize the efficiency promise of pipelined architectures faces major
challenges. On the one hand, the multi-head attention (MHA) operation in ViTs involves
computing the correlation of each token with all other tokens in an image.
This introduces global computation dependency and causes either pipeline bubbles \cite{butterfly} or high buffer cost \cite{via}.
On the other hand, pipelined architectures compute multiple
layers simultaneously and natively require more hardware resources, e.g.,
digital signal processing (DSP) blocks. The complex non-linear functions in ViT,
including GeLU, Softmax, LayerNorm, etc, also require high-precision computation
and DSP usage\cite{edge-moe, tea-s, base-2}.
As shown in Figure~\ref{fig: roofline model}, only 3.2 TOP/s can be achieved for a coarse-grained pipeline design due to the DSP limitation. If lookup tables (LUTs) are also utilized to construct the PEs, the roofline can be improved, but the design will be limited by bandwidth again and can only achieve 7.8 TOP/s.

To address the challenges, we introduce \method~in this paper. 
% We designed \method~to address the challenges.
It combines the features of fine-grained and coarse-grained pipelines to eliminate the bubbles as well as reduce buffer costs. 
\method~keeps as many weights on-chip as possible to maximize the computational intensity.
It further introduces careful approximations for activation functions to enable LUT-based processing
to combat the DSP resource limits. 
As shown in Figure~\ref{fig: roofline model}, \method~breaks through both the DSP roofline and bandwidth limitations, achieving a throughput of 17.8 TOP/s.
The contribution of \method~can be summarized as follows:

\ml{Change low-precision quantization to low bit-width quantization?}

\begin{itemize}
    \item \method~features a hybrid-grained pipelined architecture to simultaneously achieve
    low off-chip memory access, low buffer requirements, and negligible pipeline bubbles.
    \item \method~leverages \sout{low-precision}low bit-width quantization and introduces careful
    approximations for activation functions.
    The abundant LUT resources are utilized to process both linear and non-linear operators, achieving a higher roofline.
    \item \method~demonstrates 2.72$\times$ better throughput and 2.46$\times$ better
    resource efficiency over prior-art accelerators on a ZCU102 FPGA. It realizes end-to-end ViT acceleration on a single VCK190 FPGA and achieves 7118 images/s.
\end{itemize}
\section{Background}
\label{sec: background}

\subsection{Non-linear Functions in Transformers}
\label{subsec: non-linear}

% ViTs involve various non-linear functions as listed as follows:
ViTs involve various non-linear functions. Different from linear functions, it is challenging to implement them efficiently on FPGAs.

\textbf{GeLU} is applied in the MLP block and can be computed as :
\begin{equation}
  \label{eq: gelu}
    \text{GeLU}(x) = \frac{x}{2} \left(1 + \text{erf} \left(\frac{x}{\sqrt{2}}\right)\right),\text{where}\, \text{erf}(x) = \frac{2}{\sqrt{\pi}} \int_0^x \operatorname{exp}(-t^2)\text{d}t
\end{equation}

\textbf{LayerNorm}, as shown in Eq.~\ref{eq: layernorm}, is applied in both the MHA block and the MLP block.
The division and the square root operations are fused as the ``Rsqrt'' operator for hardware efficiency.
\begin{equation}
  \label{eq: layernorm}
  \begin{aligned}
  \text{LayerNorm}(\mathbf{x}) = \cfrac{\mathbf{x} - \mathrm{E}[\mathbf{x}]}{\sqrt{\mathrm{Var}(\mathbf{x})}} = ( \mathbf{x} - \mathrm{E}[\mathbf{x}] ) \cdot \operatorname{Rsqrt}(\mathrm{Var}(\mathbf{x}) )
  \end{aligned}
\end{equation}

\textbf{Softmax}, as shown in Eq.~\ref{eq: softmax}, is applied in the MHA block to normalize the attention scores.
In Softmax, the exponential function (Exp) and the reciprocal function (Recip) are required.
\begin{equation}
  \label{eq: softmax}
  \begin{aligned}
  \text{Softmax}(\mathbf{x}) = \frac{ \text{Exp} (\mathbf{x} - \mathbf{x}_\text{max} )  }{ \sum  \text{Exp}(\mathbf{x} - \mathbf{x}_\text{max}  ) }
  \end{aligned}
\end{equation}

We also regard the \textbf{ReQuant} operator in the quantized networks as a non-linear function.
ReQuant can be computed as Eq.~\ref{eq: requant}, where $\alpha_\text{int}$ and $S_\text{fixed}$ denote the integer zero point and
the fixed-point scaling factor, respectively. ReQuant requires high-precision multiplication and therefore consumes DSPs.
\begin{equation}
  \label{eq: requant}
  \begin{aligned}
  \text{ReQuant}(x) = \operatorname{clamp} \left( \left\lceil (x - \alpha_\text{int}) \cdot S_\text{fixed} \right\rfloor, Q_\text{min}, Q_\text{max} \right)
  \end{aligned}
\end{equation}

To implement these non-linear functions, three methods have been developed by prior works:
\begin{itemize}
    \item \textit{Floating point implementation} employs 32-bit or 16-bit floating point computation. Despite its simplicity and precision, it demands substantial DSPs and LUTs. 
    \item \textit{Fixed-point polynomial approximation} uses low-order polynomials to implement non-linear functions within specified ranges\cite{ibert}.
    This method is a compromise between computational complexity and accuracy.
    \item \textit{Lookup table method} involves discretizing the function input range and recording the output. While it reduces DSP usage, it usually requires Block RAMs (BRAMs) for accurate sampling\cite{edgemoe}.
\end{itemize}

\subsection{FPGA-based ViT Acceleration}
\label{subsec: stream}

\newcommand{\yes}[0]{\ding{52}}
\newcommand{\no}[0]{\ding{56}}
\renewcommand{\arraystretch}{1.0}
\definecolor{Gray}{gray}{0.85}

\begin{figure}[!tb]
    \centering
    \subfloat[The difference between pipeline paradigms: (1) temporal architecture, (2) coarse-grained pipeline architecture, (3) fine-grained pipeline architecture. "Aux" is short for auxiliary ops(non-linear functions and off-chip memory access).]{
      \includegraphics[width=0.48\textwidth]{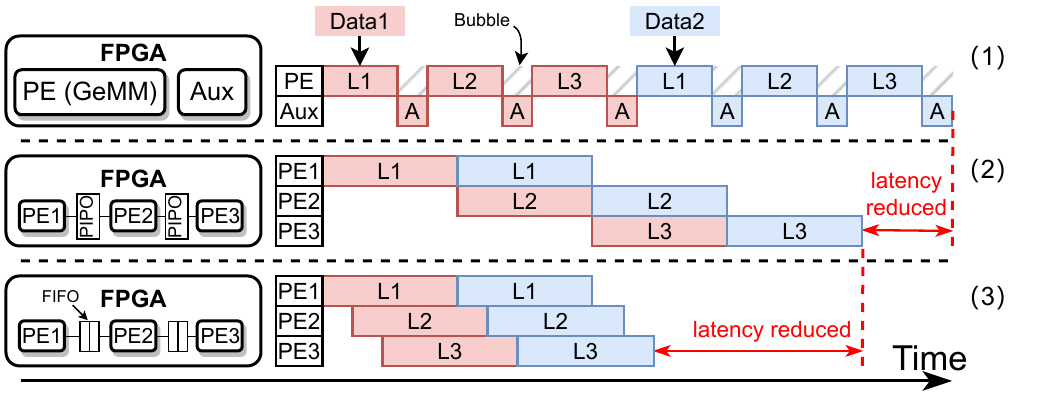}
      \label{fig: pipeline diff}
    }
    \hfill
    \subfloat[The difference between PIPO and FIFO.]{
      \includegraphics[width=0.48\textwidth]{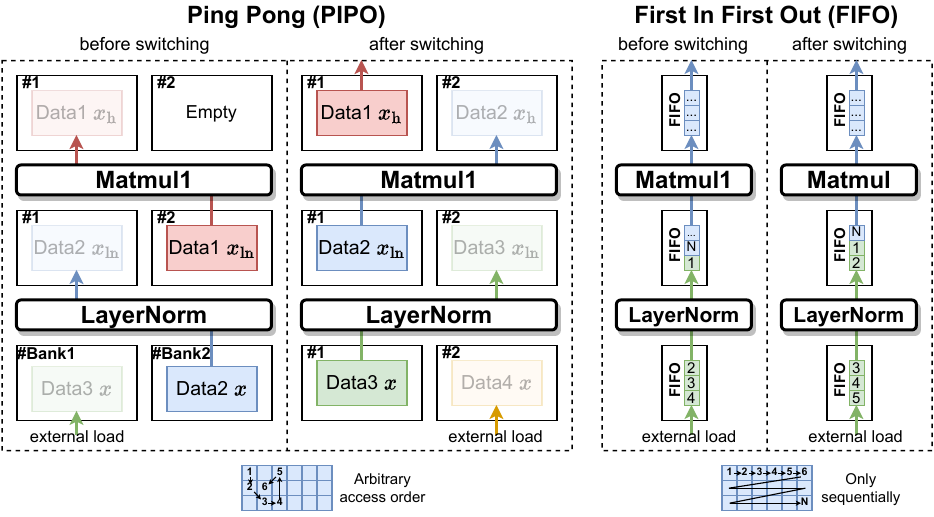}
      \label{fig: buffer diff}
    }
   
    \subfloat[Qualitative comparison between different paradigms.]{
    \scriptsize

\begin{tabular}{p{2.2cm}                 |                 p{1.0cm}                              p{1.0cm}                    p{1.0cm}            p{1.0cm}      }
\hline
\hline
                                            &{No pipeline \newline (GeMM)}        &{Coarse-grained pipeline}    &{Fine-grained pipeline}  &{Hybrid-Grained Pipeline}    \\
\hline
\textbf{Representative Work}                &\cite{high_freq_sa}                  &\cite{length_adaptive}       &\cite{DNNBuilder}        &Ours                         \\
\hline
\textbf{Buffer Type}                        &Global Buffer                        &PIPO                         &FIFO                     &Buffer +\newline FIFO        \\
\hline
\textbf{Buffer Cost}                        &\green{Small}                        &\red{Large}                  &\green{Small}            &\blue{Mid}                   \\
\hline
\textbf{Data Access Order}                  &\green{Any order}                    &\green{Any order}            &\red{Sequentially}       &\green{Any order}            \\
\hline
\textbf{Data Access Times}                  &\green{Multiple}                     &\green{Multiple}            &\red{Only Once}           &\green{Multiple}             \\
\hline
\textbf{ViT Compatibility}                  &\green{\yes}                         &\green{\yes}                 &\red{\no}                &\green{\yes}                 \\
\hline
\textbf{Throughput}                         &\red{Low}                            &\green{High}                 &\green{High}             &\green{High}                 \\
\hline
\textbf{Latency}                            &\red{High}                           &\blue{Mid}                   &\green{Low}              &\green{Low}                  \\
\hline
\hline
\end{tabular}
    }
     \caption{Compare coarse-grained and fine-grained pipeline.}
    \label{fig: compare_fifo_pipo}
    \Description{Compare coarse-grained and fine-grained pipeline}
    
\end{figure}

The FPGA-based ViT accelerators can be categorized into two architectures: temporal architecture and pipelined architecture. As shown in Figure ~\subref*{fig: pipeline diff}, temporal architectures leverage unified PEs to process different layers.
Such PE is typically dedicated to performing General Matrix Multiplication, commonly referred to as GeMM.
Most existing FPGA-based ViT accelerators belong to the category
\cite{high_freq_sa, heat-vit, HPTA, group_vector_sa, vis-top, peeling, auto-vit-acc, swin_accelerator, calabash, trac, edge-moe, Saust, VAQF, accel-tran, EFA-trans, vit-cod}.
Efficient systolic arrays \cite{sa} or sparsity-aware PEs \cite{heat-vit, vit-cod} enable temporal architectures to accelerate linear layers effectively. 
However, these architectures often necessitate frequent off-chip memory access for intermediate results and tend to underutilize resources due to a lack of concurrent multi-operator execution.

Pipelined architectures, as illustrated in Figure ~\subref*{fig: pipeline diff}, optimize layer processing by customizing PEs for varied operators across layers \cite{via, butterfly}, enhancing resource utilization and minimizing off-chip memory costs through inter-PE data transfer \cite{via, TranCIM, p3vit}. Coarse-grained pipeline processes entire tensors for different operators \cite{quant_bert, unified_attn_conv, e2e_speech, p3vit, sparsity-multi-heads, length_adaptive, via}. On the contrary, fine-grained pipeline tiles activation tensors for sub-tensor processing\cite{FixyFPGA, tomato}. A coarse-grained pipeline uses Ping Pong (PIPO) buffers, requiring double the tensor’s memory size, while a fine-grained pipeline employs First In First Out (FIFO) buffers with tile-level granularity. The behavior of PIPO and FIFO are depicted in Figure ~\subref*{fig: buffer diff}. Fine-grained pipelined architectures are predominantly utilized in CNN acceleration, providing high hardware utilization and low buffer costs \cite{HPIPE, tomato, FixyFPGA, GoingDeeper}, whereas coarse-grained architectures are preferred for ViT acceleration \cite{quant_bert, unified_attn_conv, e2e_speech, p3vit, sparsity-multi-heads, length_adaptive, via} for data access ability. 
A comparative analysis of these architectures is presented in Table~\ref{fig: compare_fifo_pipo}(c).
\section{Challenges}
\label{sec: challenges}

\begin{figure}[htb]
  \centering
  \includegraphics[width=0.48\textwidth]{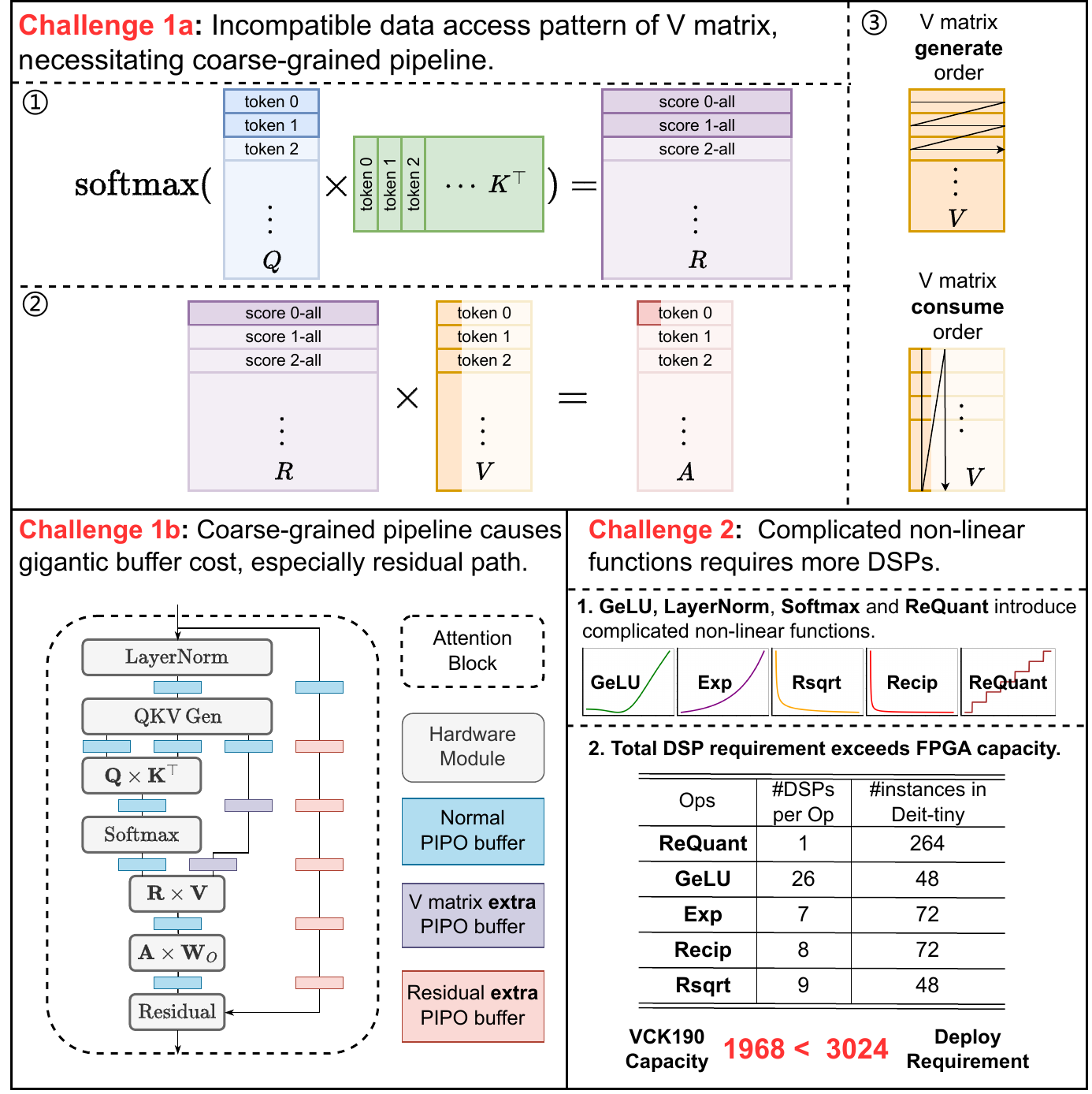}
  \caption{The challenges of ViT hardware acceleration.}
  \label{fig: challenges}
  \Description{The challenges of ViT hardware acceleration.}
\end{figure}

Although ViT has outstanding performance, its complexity brings challenges for FPGA accelerator design as depicted in Figure \ref{fig: challenges}.
These challenges motivated us to propose HG-PIPE.

\paragraph{Challenge 1: Dataflow Design Dilemma}
\label{subsec: challenge 1}

The self-attention mechanism in ViTs introduces global computation dependencies, 
leading to challenges in data locality and dataflow design. 
As illustrated in Figure \ref{fig: challenges} (1a), the transpose operation within the self-attention block breaks the data access pattern for the $\mathbf{V}$ matrix, hampering the continuity needed for fine-grained pipelines and necessitating coarse-grained pipelines.
However, as shown in Figure \ref{fig: challenges} (1b), the coarse-grained approach causes gigantic activation buffering costs, exacerbated by residual connections that require multiple Ping Pong buffers to avoid deadlock. 
Specifically, in the Deit-tiny network, buffering one residual tensor consumes 14 BRAMs. One attention block in a coarse-grained pipeline requires 6 PIPO stages (168 BRAMs) just for the residual path.
This extensive buffering demand is impractical for FPGA platforms\gqy{\cite{via}}\cite{via}. \ml{Add a reference here.} 

\paragraph{Challenge 2: High DSP Usage}
\label{subsec: challenge 2}

The extensive non-linear functions in ViT present challenges for efficient FPGA implementation.
In our HLS synthesis experiments, naive floating-point implementations of functions like \textbf{Exp}, \textbf{Rsqrt}, and \textbf{Recip} are DSP-intensive, consuming 7, 8, and 9 DSPs respectively. The \textbf{GeLU} function is even more DSP-intensive, requiring 26 DSPs. The \textbf{ReQuant} function additionally uses 1 DSP. In the estimation, implementing these non-linear functions in a Deit-tiny model requires 3024 DSPs, exceeding the DSP capacity of a VCK190 FPGA.
\ml{This is a bit weird.}
\sout{On the edge side, processes like Fast Fourier Transform (FFT) and Finite Impulse Response (FIR) filters may also require large amounts of DSPs. }
Therefore, reducing DSP usage in ViT accelerators is crucial for FPGA implementation.
\section{Main Methods}

\subsection{Overview of HG-PIPE}
\label{subsec: methodology}

\begin{figure}[!th]
\centering
\includegraphics[width=0.45\textwidth]{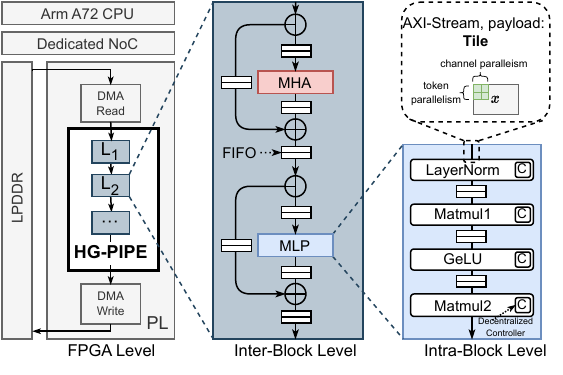}
\caption{Overview of the HG-PIPE accelerator design.}
\label{fig: overview}
\Description{Overview of the HG-PIPE accelerator design.}
\end{figure}

\begin{comment}
\begin{figure}[!tb]
\centering
\includegraphics[width=0.5\textwidth]{figs/attention_module.pdf}
\caption{The dataflow design of the MHA module.}
\label{fig: attention_module}
\Description{The dataflow design of the MHA module.}
\end{figure}
\end{comment}

\begin{figure*}[!tb]
\centering
\includegraphics[width=1\textwidth]{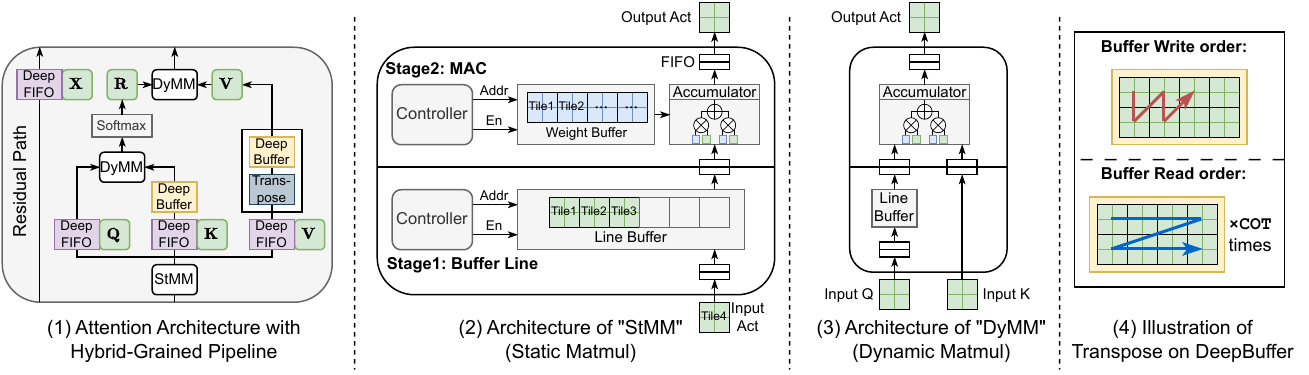}
\caption{The dataflow design of the MHA module.}
\label{fig: attention_module}
\Description{The dataflow design of the MHA module.}
\end{figure*}

\begin{figure}[!tb]
\centering
\includegraphics[width=0.5\textwidth]{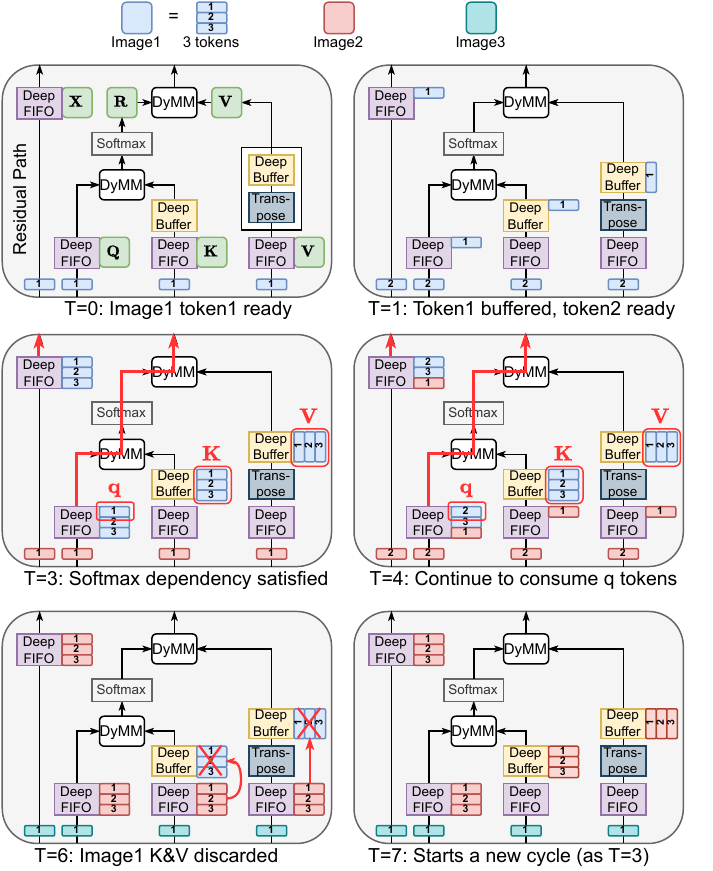}
\caption{The time diagram of the hybrid-grained pipeline.}
\label{fig: hybrid_grained}
\Description{The time diagram of the hybrid-grained pipeline.}
\end{figure}

\begin{figure}[!tb]
\centering
\includegraphics[width=0.5\textwidth]{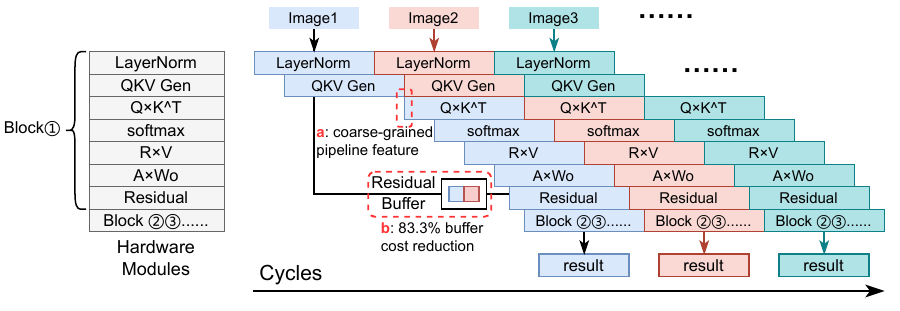}
\caption{The hardware view of the hybrid-grained pipeline.}
\label{fig: hardware view}
\Description{The hardware view of the hybrid-grained pipeline.}
\end{figure}

The overview of the HG-PIPE accelerator is depicted in Figure \ref{fig: overview}. It is integrated into an FPGA System-on-Chip (SoC) framework, with multiple peripheral components, including external memory, host system, and network-on-chip (NoC).

HG-PIPE stands out with its fully pipelined architecture across all layers, eliminating external memory access for intermediate activations or weights. It directly processes incoming data from the Direct Memory Access (DMA) module, channeling the final output back to external memory. HG-PIPE utilizes dedicated modules for each component, avoiding time division multiplexing. This design principle extends from the Inter-Block to the Intra-Block Level.

To achieve sufficient throughput, HG-PIPE incorporates over 20,000 MAC units. Managing such a vast array of MAC units typically complicates control logic. 
Centralized control often leads to extended routing nets that hamper timing. As seen in FixyFPGA~\cite{FixyFPGA}, the accelerator utilizes over 1336k LUTs and only reaches a 132MHz clock frequency.
To mitigate this, HG-PIPE adopts an asynchronous, decentralized pipeline strategy, where each stage is controlled by its own FSM. With handshakes on the AXI-Stream interface, modules are completely decoupled.
The design incorporates FIFOs within these connections to cover data generation fluctuations, avoiding deadlocks in the data flow. As illustrated, data is transferred in the form of tiled tensors.

\subsection{Hybrid-Grained Pipeline}
\label{subsubsec: hybrid-grained pipeline}

To address the challenges outlined in Sec. \ref{subsec: challenge 1}, we developed a hybrid-grained pipeline to effectively manage computing and buffering across different granularities. Despite the dependencies across tokens introduced by the Softmax operator and transposed matrix multiplication, the attention mechanism retains substantial locality in other segments. Notably, the attention block comprises four branches: the residual branch, and Q/K/V branches. The Q branch exhibits good data locality, as illustrated in Figure \ref{fig: challenges} (1a), where generating one $\mathbf{A}$ token requires one $\mathbf{Q}$ token and the entirety of $\mathbf{K}$ and $\mathbf{V}$ tokens. In particular, the V matrix requires row-wise access, differing from token-wise access.

To solve the granularity problem, we implemented deep buffers in the K and V branches and introduced a "Transpose Module" in the V branch to re-order access, aligning it with the Q matrix's fine-grained pipeline, depicted in Figure \ref{fig: attention_module} (1) and (4). The buffer is deep enough to hold the entire $\mathbf{K}$ or $\mathbf{V}$ tensor. The content of K and V buffers will be read $\mathtt{COT}$ times to generate all output channels, which will be explained in Sec. \ref{subsec: parallelism design}. Figure \ref{fig: attention_module} (2) and (3) illustrate the architecture of "StMM" and "DyMM". "StMM" represents matrix multiplications (MMs) with static weights (e.g., QKV generation and MLP MMs), and "DyMM" represents MMs with dynamic weights (e.g., $\mathbf{Q}\times\mathbf{K}^\top$ and $\mathbf{R}\times\mathbf{V}$). 
They both consist of two stages, with the first stage buffering all input channels and the second stage performing MAC operations. The difference is that "DyMM" streams the dynamic weight from the previous stage buffer while "StMM" freezes the weights as ROMs. The deep buffers guarantee coarse-grained access for dynamic weights (i.e., $\mathbf{K}$ and $\mathbf{V}$).

However, deep buffers alone were insufficient since other branches would be halted in the execution.
Thus, we incorporated deep FIFOs into all four branches to maintain continuous execution.
We carried out simulation experiments to identify the shallowest depth that avoids deadlocks, and the typical depth of deep FIFOs is 512.
The mechanism of deep FIFOs and deep buffers is illustrated in Figure \ref{fig: hybrid_grained}. The process is simplified and focuses on key stages for clarity, with each image containing 3 tokens.
At T=0, the pipeline is initialized with the first token of Image1 (blue) queued in the prior stage. The residual and Q tokens are stored in deep FIFOs, while K and V tokens are placed directly into deep buffers, with V tokens being transposed first. By T=3, all K and V tokens of Image1 are buffered, therefore the data dependencies for attention are met, allowing the first Q token to be consumed and the first A token to be generated (as highlighted by the red arrow). As Image1's Q tokens are consumed, tokens from Image2 (red) start to load. By T=6, after all Q tokens of Image1 are processed, the K and V buffers refresh for Image2, initiating a new cycle at T=7.

Figure \ref{fig: hybrid_grained} shows token-level dataflow, while our design operates at a more fine-grained sub-token granularity, detailed in Sec. \ref{subsec: parallelism design}. Figure \ref{fig: hardware view} provides a hardware perspective, demonstrating overlaps of operators and hybrid-grained features. "QKV Gen" module initiates before "LayerNorm" completes as a fine-grained pipeline, and as highlighted (\red{a}), "$Q\times K^\top$" cannot start early due to dependencies as a coarse-grained pipeline. The residual buffer cost is significantly reduced by 83.3\% compared to traditional PIPO implementation (\red{b}).

\subsection{Parallelism Design} 
\label{subsec: parallelism design}

\begin{figure}[!tb]
\centering
\begin{minipage}{0.13\textwidth}
  \centering
  \includegraphics[width=\linewidth]{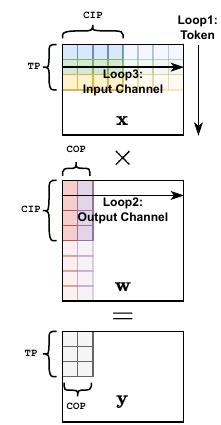}
\end{minipage}
\begin{minipage}{0.34\textwidth}
  \centering
  \renewcommand{\arraystretch}{1.2}
  \scriptsize
  \begin{tabular}{|p{1.2cm}|m{3.6cm}|}
    \hline
    \textbf{Notation}                                   & \textbf{Explanation} \\
    \hline
    $\mathtt{T}$                                        & Sequence length \\
    $\mathtt{CI}$                                       & Embedding dimension (input channels) \\
    $\mathtt{CO}$                                       & Hidden layer dimension (output channels) \\
    $\mathtt{TP}$                                       & Parallelism on Sequence length \\
    $\mathtt{CIP}$                                      & Parallelism on input channels \\
    $\mathtt{COP}$                                      & Parallelism on output channels \\
    $\mathtt{TT}$                                       & Trip count on $\mathtt{T}$ direction;  $\mathtt{TT}  = {\mathtt{T}} /{\mathtt{TP}}$ \\
    $\mathtt{CIT}$                                      & Trip count on $\mathtt{CI}$ direction; $\mathtt{CIT} = {\mathtt{CI}}/{\mathtt{CIP}}$ \\
    $\mathtt{COT}$                                      & Trip count on $\mathtt{CO}$ direction; $\mathtt{COT} = {\mathtt{CO}}/{\mathtt{COP}}$ \\
    $\mathbf{x, w, y}$                                  & Input tokens, weight, output tokens, \newline $\mathbf{x}$.shape = ($\mathtt{T}$, $\mathtt{CI}$), $\mathbf{w}$.shape = ($\mathtt{CO}$, $\mathtt{CI}$), $\mathbf{y}$.shape = ($\mathtt{T}$, $\mathtt{CO}$) \\
    $\text{DW}_\mathbf{x}, \text{DW}_\mathbf{w}$        & Data width of $\mathbf{x}$, $\mathbf{w}$ \\
    $\text{B}_\text{width}, \text{B}_\text{depth}$      & BRAM bank width, depth \\
    \hline
  \end{tabular}
\end{minipage}
\caption{The notations of tiled matrix multiplication.}
\label{fig: tiled matrix multiplication}
\Description{The tiled matrix multiplication.}
\end{figure}

Matrix Multiplication (MM) is the primary operator used in ViT computations. 
We implement tiled MM using an Output Stationary (OS) dataflow to minimize partial sum storage costs, as illustrated in Figure \ref{fig: tiled matrix multiplication}. 
MM involves three nested loops: Token, Output Channel, and Input Channel. All loops are tiled to enhance data locality.
For clarity, notations are detailed in the Table.
We meticulously design the parallelism of all the modules for two reasons: pipeline balance and BRAM utilization efficiency.\footnote{We didn't apply an automatic process for generating parallelism hyperparameters, since in transformers, each layer has the same structure and shape, deriving 
a small design space. Hand-crafted design is feasible and good enough.}

\begin{figure}[!tb]
    \centering
    \subfloat[Eliminating imbalance-induced bubble in a fine-grained pipeline.]{%
        \includegraphics[width=0.45\textwidth]{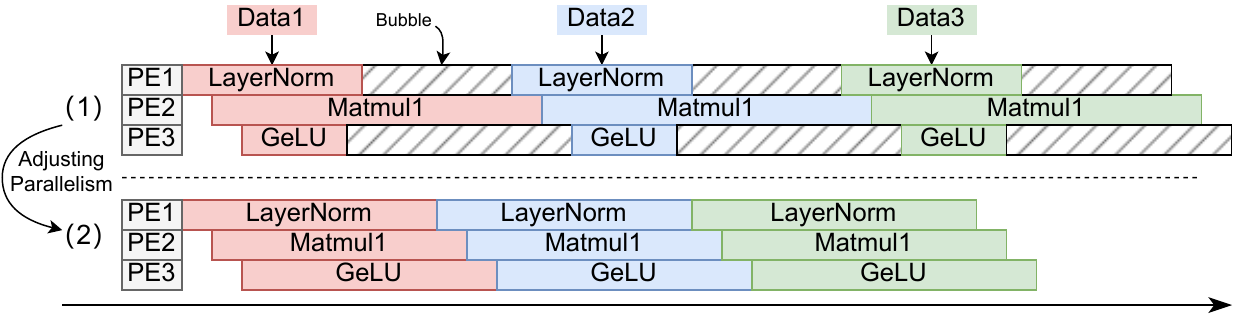}
        \label{fig: balance bubble}
    }\\
    \subfloat[Adjust the parallelism design to improve BRAM utilization.]{%
        \includegraphics[width=0.45\textwidth]{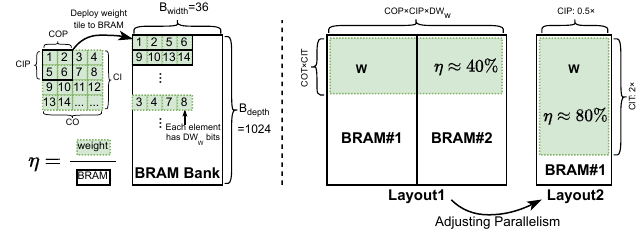}
        \label{fig: improper memory layout}
    }
    \caption{Adjusting parallelism to improve efficiency.}
\end{figure}

\subsubsection{Pipeline Balance}
The pipeline balance is essential for the accelerator's performance. 
The Initiation Interval (II) of the whole pipeline is the maximum of the II of all the pipeline stages. As shown in Figure \ref{fig: balance bubble}(1), if the cycles are not balanced, the bubbles will be generated.
Therefore, by allocating more computing resources to the Matmul1 module, the pipeline can be balanced in Figure \ref{fig: balance bubble}(2).
By adjusting the parallelism, we try to make the II of the pipeline stages balanced. The calculation of II is detailed in Table \ref{tab: parallelism design} footnotes.

\subsubsection{BRAM Utilization Efficiency}
BRAM is a scarce resource on FPGA for AI applications. 
It is also a critical resource for our design since HG-PIPE needs to
freeze as much of the weights on-chip as possible to reduce external memory access.
The parallelism design can directly decide the memory layout of the weights in BRAM (as shown in Figure \ref{fig: improper memory layout}), and improper memory layout will lead to BRAM waste. In Layout1, the weight $\mathbf{w}$ requires two BRAMs to accommodate it. However, by scaling $\mathtt{CIP}$ to half of its original value in Layout2, there will only one BRAM be required. As illustrated in Table \ref{tab: parallelism design} footnotes, we can calculate the number of BRAMs required by one MM module (denoted as $\#\text{BRAM}$), and the corresponding BRAM utilization efficiency (denoted as $\eta$).

\subsubsection{Parallelism Design Results}

\renewcommand{\arraystretch}{1.2}
\begin{table}[!tb]

\centering
\begin{threeparttable}
\tiny
% \scriptsize
% \footnotesize
\begin{tabular}{|ll|l|l|l|l|l|l|l|}
\hline
\hline
\multicolumn{2}{|c|}{Module Name}                                                       & \texttt{T/TP=TT}      & \texttt{CI/CIP=CIT}       & \texttt{CO/COP=COT}   & MOPs\tnote{1} & P\tnote{2}    & II\tnote{3}       & $\eta$\tnote{4}       \\ \hline
\multicolumn{1}{|l|}{\multirow{7}{*}{MHA}}                      & LayerNorm             & 196/2=98              & 192/1=192                 & -                     & 0.11          & 2             & 56448                 & -                     \\ \cline{2-9} 
\multicolumn{1}{|l|}{}                                          & QKV Gen               & 196/2=98              & 192/6=32                  & 64/4=16               & 2.41          & 48            & 50176                 & 100\%                 \\ \cline{2-9} 
\multicolumn{1}{|l|}{}                                          & QK MatMul             & 196/2=98              & 64/4=16                   & 196/7=28              & 2.46          & 56            & 43904                 & 68.1\%                \\ \cline{2-9} 
\multicolumn{1}{|l|}{}                                          & Softmax               & 196/2=98              & 196/1=196                 & -                     & 0.11          & 2             & 57624                 & -                     \\ \cline{2-9} 
\multicolumn{1}{|l|}{}                                          & RV MatMul             & 196/2=98              & 196/7=28                  & 64/4=16               & 2.46          & 56            & 43904                 & 68.1\%                \\ \cline{2-9} 
\multicolumn{1}{|l|}{}                                          & Output Proj           & 196/2=98              & 192/12=16                 & 192/6=32              & 7.23          & 144           & 50176                 & 100\%                 \\ \cline{2-9} 
\multicolumn{1}{|l|}{}                                          & Residual Add          & 196/2=98              & 192/1=192                 & -                     & 0.038         & 2             & 18816                 & -                     \\ \hline
\multicolumn{1}{|l|}{\multirow{4}{*}{MLP}}                      & LayerNorm             & 196/2=98              & 192/1=192                 & -                     & 0.11          & 2             & 56448                 & -                     \\ \cline{2-9} 
\multicolumn{1}{|l|}{}                                          & MatMul1               & 196/2=98              & 192/12=16                 & 768/24=32             & 28.9          & 576           & 50176                 & 100\%                 \\ \cline{2-9} 
\multicolumn{1}{|l|}{}                                          & GeLU                  & 196/2=98              & 192/2=98                  & -                     & 0.15          & 4             & 37632                 & -                     \\ \cline{2-9} 
\multicolumn{1}{|l|}{}                                          & MatMul2               & 196/2=98              & 768/24=32                 & 192/12=16             & 28.9          & 576           & 50176                 & 100\%                 \\ \hline\hline
\end{tabular}
\begin{tablenotes}
    \item[1] MOPs: Million Operations. $\text{MOPs} = \mathtt{T} \times \mathtt{CI} \times \mathtt{CO}$.
    \item[2] P: Parallelism in total. For Matrix Multiplication, P is the number of parallel MAC units. For other modules, P is the number of parallel elementwise or reduction units. $\mathtt{P} = \mathtt{TP} \times \mathtt{CIP} \times \mathtt{COP}$.
    \item[3] II: Initiation Interval, which means the number of cycles required by the module for one inference. For \textbf{LayerNorm} and \textbf{Softmax} modules, three passes are needed, therefore they require 3 times cycles. $\text{II} = \mathtt{TT} \times \mathtt{CIT} \times \mathtt{COT}$. For an accelerator with multiple stages, the II of the whole accelerator is the maximum of the II of all the stages: $\text{II}_{\text{accelerator}} = \max(\text{II}_{\text{stage1}}, \text{II}_{\text{stage2}}, \cdots, \text{II}_{\text{stageN}})$.
    \item[4] $\eta$: BRAM Efficiency. 
    % Detailed computation is shown in Figure \ref{fig: improper memory layout}. 
    $\text{\#BRAM} = \left\lceil \cfrac{\text{DW}_\mathbf{w} \cdot \mathtt{CIP} \cdot \mathtt{COP}} { \text{B}_\text{width}} \right\rceil \cdot \left\lceil \cfrac{\mathtt{CIT} \cdot \mathtt{COT}} { \text{B}_\text{depth}} \right\rceil$, and $ \eta = \cfrac{\text{DW}_\mathbf{w}\cdot \mathtt{CI} \cdot \mathtt{CO} }{ \text{\#BRAM} \cdot \text{B}_\text{width} \cdot \text{B}_\text{depth} }$.
\end{tablenotes}
\caption{The parallelism design result on Deit-tiny.}
\label{tab: parallelism design}
\end{threeparttable}
\end{table}

The parallelism design results are shown in Table \ref{tab: parallelism design}, in which we tried to achieve the two goals at the same time.
As demonstrated, most modules have closely matched IIs. To save DSP resources, we choose the non-linear operators to be the II bottleneck (57624 of the Softmax module). For the Residual Add module, we keep $\mathtt{TP}=2$ to simplify the design, leading to smaller II values. Although bubbles exist in Residual, this is not a big waste since it only requires 0.038 MOPs. For the BRAM utilization efficiency, we achieved 100\% efficiency for all the modules that employ static weight (QKV Gen, Output Proj in MHA block and MatMul1, MatMul2 in MLP block).

\subsection{Efficient and Accurate LUT-based Processing}
\label{subsec: approximate non-linear functions}
ViTs incorporate complex non-linear functions, previously discussed in Sec. \ref{subsec: non-linear} and Sec. \ref{sec: challenges}. Accurately computing these functions is crucial for maintaining model performance, but full-precision floating-point calculations exceed FPGA DSP capacity. In this section, we provide optimization and approximation strategies that enhance hardware friendliness while preserving accuracy, allowing non-linear functions to be effectively implemented using LUTs.

\subsubsection{LUT-based MAC unit}
The implementation of multiplication operations using LUTs is a technique in FPGA-based computations. Consider a scenario where operands of multiplication are quantized to 3 bits. In this case, the multiplication operation can be decomposed into six boolean functions, with each function consuming 6 bits to produce a single bit of the multiplication. As a result, only 6 LUT-6 are required. The adoption of LUTs for MAC units can significantly enhance the computational capabilities of an FPGA, effectively raising its computation roofline. This approach has been successfully utilized in several accelerator works~\cite{auto-vit-acc, FixyFPGA, LUTNet, tomato}. We have incorporated this technique in our design as well.

\subsubsection{Power-of-Two Index Approximation}
\label{subsubsec: pot}
In the LUT method, the index computation is essential. The process is similar to quantization in Eq. \ref{eq: requant}: discretizing continuously distributed input values.
In the traditional table method, the index is computed as Eq. \ref{eq: index computation}. 
\begin{equation}
    \text{index} = \left\lceil (\text{data} - \alpha) \times \cfrac{ 2^n - 1 } { \beta - \alpha } \right\rfloor
    \label{eq: index computation}
\end{equation} in which $(\alpha, \beta)$ is the data range of input tensor, and $n$ is the address width of the table.
% The formation of Eq. \ref{eq: index computation} is very similar to Eq. \ref{eq: requant}.
However, this method will require a DSP for the multiplication. This is contradictory to the motivation of the table method: reduce DSP usage.
The similarity between index computation and quantization inspired us to introduce a quantization technique into our LUT methods: Power-of-Two Quantization ~\cite{Logarithm, MixMatch, tomato}.
This method estimates the scaling factor with its nearest Power-of-Two (PoT) value, therefore simplifying the high-precision multiplication to a bit shifting. The estimation is shown in Eq. \ref{eq: pot index}.

\begin{equation}
    \text{index}_\text{PoT} = (\text{data} - \alpha) \gg s_\text{PoT}, \quad \text{where } s_\text{PoT} = \left\lceil \log_2 \left( \cfrac{\beta - \alpha}{2^n-1} \right) \right\rceil
    \label{eq: pot index}
\end{equation}

\begin{figure}[!tb]
    \centering
    \subfloat[The computation of index and the Power-of-Two approximation method.]{
        \includegraphics[width=0.5\textwidth]{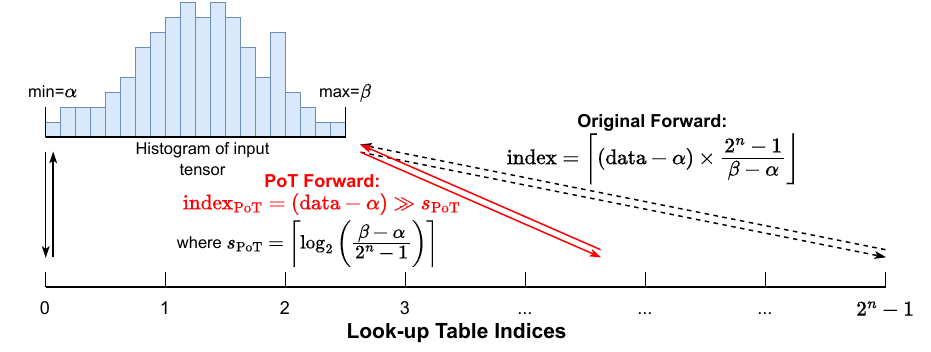}
        \label{fig: index computation}
    }
    \\
    \subfloat[The fusion of GeLU operator and ReQuant operator.]{
        \includegraphics[width=0.5\textwidth]{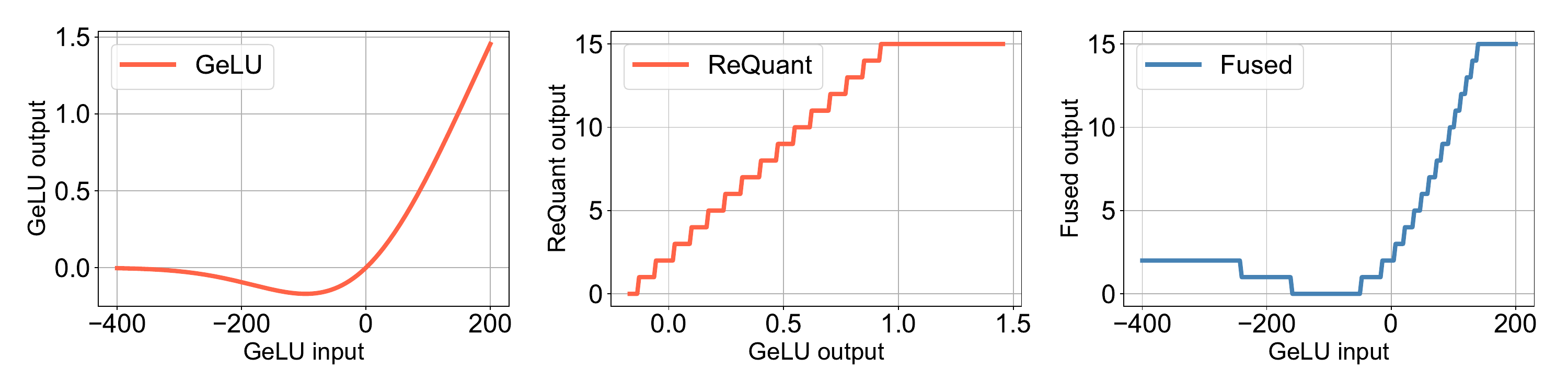}
        \label{fig: gelu_fusion}
    }
    \\
    \subfloat[The joint table range calibration for GeLU and ReQuant.]{
        \includegraphics[width=0.5\textwidth]{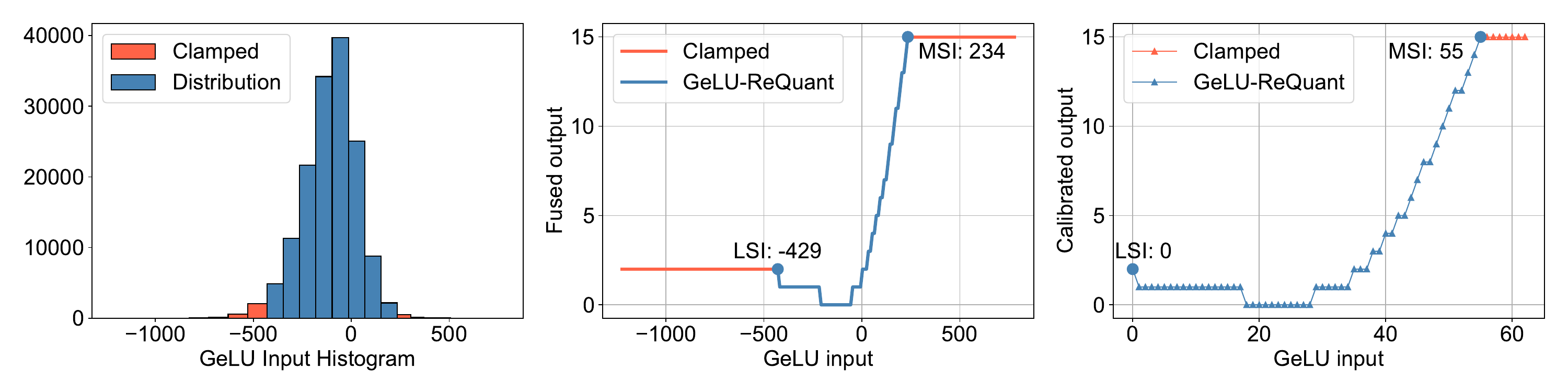}
        \label{fig: joint calibration}
    }
    \\
    \subfloat[The segmented table for the Recip operator with high dynamic range. The segmentation pivot is annotated on the curve.]{
        \includegraphics[width=0.45\textwidth]{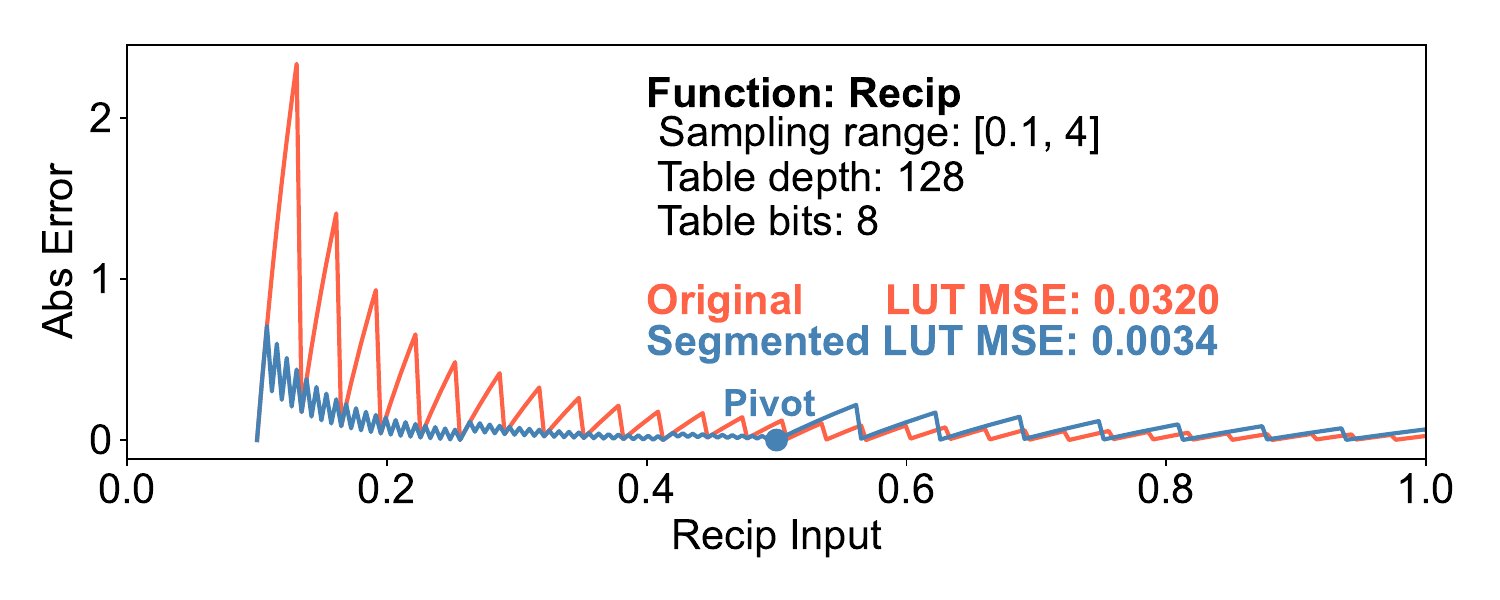}
        \label{fig: segmented table}
    }
    \caption{LUT optimization techniques.}
\end{figure}

Our estimation is a bit different from vanilla PoT Quantization. We apply a ceiling instead of rounding to avoid index overflowing. The whole process is shown in Figure \ref{fig: index computation}. The blue bars show the data distribution of the input tensor. Taking the min and max values, the normal scaling factor can be calculated, and the black dashed arrows show the normal index mapping, aligning the boundaries and mapping $\beta$ to the highest index ($2^n-1)$. On the contrary, as shown by the red arrows, PoT estimation will not guarantee boundary alignment, but it will make sure the scaling happens as a static bit shifting.

\subsubsection{GeLU-ReQuant Operator Fusion}
In a quantized network, quantizers are inserted before all MMs to improve MAC efficiency. Fusing quantizers with preceding operators simplifies the logic and reduces LUT consumption. The fusion involves sampling a combined transfer curve, depicted in Figure \ref{fig: gelu_fusion}. The left sub-figure represents the GeLU function, the middle one the ReQuant function with 4-bit precision, and the right one the resulting fused curve. The sampling will happen on the combined curve.

\subsubsection{Implement ReQuant as Table}
Besides GeLU fusions, numerous ReQuant operators cannot be fused and also consume DSP resources. By treating quantizers as non-linear functions, we can apply Power-of-Two index approximation to eliminate DSP usage as well. 
Employing low-precision quantization significantly reduces LUT cost. 
Our experiments show that a 64-entry ReQuant table sufficiently preserves accuracy. 

\subsubsection{Joint Table Range Calibration}
\label{subsubsec: joint table range calibration}
In the implementation result of the ReQuant table and GeLU-ReQuant table, we discovered that there are many repeated entries generated from the clamping behavior (Eq. \ref{eq: requant}) on the two ends, causing wastes of representative ability. This is demonstrated in the middle sub-figure of Figure \ref{fig: joint calibration}, in which the repeated entries on the two ends are colored orange. We introduce Joint Table Range Calibration to reduce redundancy in the table. This method aligns the input data distribution with the table contents to optimize implementation. It iteratively identifies the Least Significant Index (LSI) and Most Significant Index (MSI) to recalculate the data range and updates the table until the data range stabilizes. In Figure \ref{fig: joint calibration}, the left sub-figure shows clamped data in orange.
After calibration, the LSI is mapped to 0, and MSI is mapped near the right boundary.
There will still be a few remaining repeated entries on the right side due to PoT approximation.

\subsubsection{Segmented Table for High Dynamic Range Recip}
The Recip function in the Softmax module exhibits a high dynamic range in our statistic experiments.
To accurately sample the function results on the sheer range between 0 to 1, the table needs to be very large.
Initially, storing the reciprocal function would have needed an entire BRAM bank (depth=1024, width=36) to maintain accuracy. To minimize BRAM usage, we exploited the function's inherent properties and segmented it into two parts, each owning an independent scaling factor. 
We empirically divide the input range at the first 1/8 for the steep part and the remainder for the flat. 
This approach is visualized in Figure \ref{fig: segmented table}. The orange line is the abs error of the original LUT implementation of Recip, and the blue one is the segmented implementation with the segmentation pivot annotated to it. With more entries between 0 and 1, the sampling is more accurate, reducing Mean Squared Error (MSE) from 0.032 to 0.0034.

\subsubsection{Inversed Exponential Table}
In our experiment, we observed that the PoT approximation on Exp will cause huge accuracy degradation.
The possible explanation is that in the calculation of Softmax, each element is subtracted by the maximum value in its group to maintain numerical stability. 
Therefore, the max value of the input range is anchored to 0.
However, the approximation described in Sec. \ref{subsubsec: pot} takes the min value as the zero point. The moving of the anchor point may cause an inaccurate approximation of the sensitive anchor values. The solution is simple: we make $\beta$ the zero point and modify the index computation from Eq. \ref{eq: pot index} to Eq. \ref{eq: pot index exp}:
\begin{equation}
    \text{index}_\text{PoT} = (\beta -\text{data}) \gg s_\text{PoT}, \quad \text{where } s_\text{PoT} = \left\lceil \log_2 \left ( \cfrac{\beta - \alpha}{2^n-1} \right) \right\rceil
    \label{eq: pot index exp}
\end{equation}

\section{Experiments}
\label{sec: experiments}

\subsection{Experiment Setup}
To evaluate our design, we selected the Deit-tiny model and Deit-small model, following our baselines AutoViTAcc \cite{auto-vit-acc}, HeatViT \cite{heat-vit} and SSR\cite{ssr}. We tested on two FPGA platforms: ZCU102 and VCK190. The ZCU102 allows direct comparisons with prior works, while the VCK190 supports full deployment of the whole network. 
Throughput was measured using the PYNQ framework,
and power consumption was assessed with Xilinx's BEAM tool.

\begin{figure}[!tb]
    \centering
    \subfloat[DSP usage and accuracy with all the optimization applied step by step]{
      \includegraphics[width=0.48\textwidth]{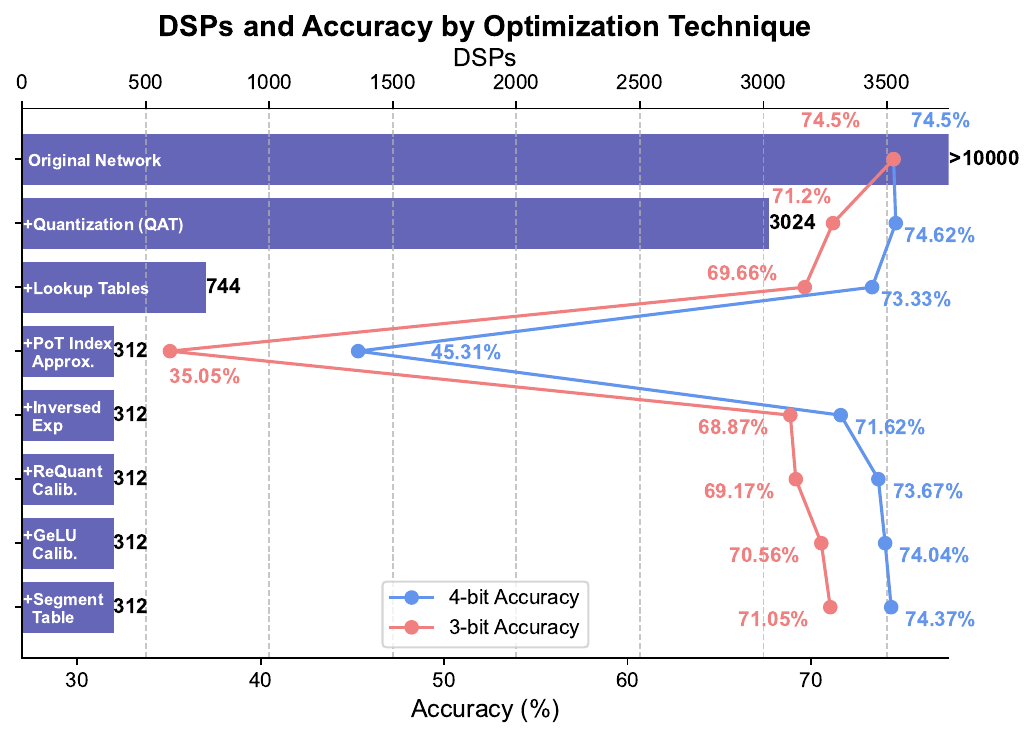}
      \label{fig: ablation_study_curve}
    }\\
    \subfloat[Ablation study experiment results.]{
    \scriptsize
\begin{tabular}{p{2.3cm} | p{1.8cm} | p{1.8cm}}
\hline
\hline
Ablation & \textbf{Deit-tiny 3bit} & \textbf{Deit-tiny 4bit} \\
\hline
w/o Inverted Exp        & 28.80\% (\textbf{-42.25\%}) & 48.87\% (\textbf{-25.50\%}) \\
w/o ReQuant Calib.      & 70.56\% (\textbf{-0.49\%})  & 74.08\% (\textbf{-0.29\%}) \\
w/o GeLU Calib.         & 69.49\% (\textbf{-1.56\%})  & 72.44\% (\textbf{-1.93\%}) \\
w/o Segmented Recip     & 70.57\% (\textbf{-0.48\%})  & 74.04\% (\textbf{-0.33\%}) \\
\hline
\hline
\end{tabular}
    \label{tab: ablation}
    }\\
    \subfloat[Resource consumption reduction with LUT methods.]{
\scriptsize
\renewcommand{\arraystretch}{1.0}
\begin{tabular}{p{1.3cm}        |  p{0.5cm} | p{0.5cm}          |  *{3}{p{0.2cm}} | *{3}{p{0.2cm}}       }
\hline
\hline
Non-linear Functions   &Table depth    &Table bits    &\multicolumn{3}{p{1.0cm}|}{LUT-6 Cost Reduction}       &\multicolumn{3}{p{1.0cm}}{DSP Cost Reduction} \\
\hline
\textbf{Exp}           &64 &8          &945         &$\rightarrow$   &50                         &7            &$\rightarrow$          &0 \\
\textbf{GeLU}          &64 &3          &1650        &$\rightarrow$   &43                         &26           &$\rightarrow$          &0 \\
\textbf{Recip}         &64*2&8         &196         &$\rightarrow$   &72                         &8            &$\rightarrow$          &0 \\
\textbf{Rsqrt}         &64 &12         &425         &$\rightarrow$   &48                         &9            &$\rightarrow$          &0 \\
\textbf{ReQuant}       &64 &3          &0           &$\rightarrow$   &3                          &1            &$\rightarrow$          &0 \\
\hline
\hline
\end{tabular}
    \label{tab: resource reduction}
    }
    \caption{Experiment results on LUT optimizations.}
    \label{fig: luts}
    \Description{Experiment results on LUT optimizations}
    
\end{figure}

\begin{figure}[!tb]
\centering
    \includegraphics[width=1\linewidth]{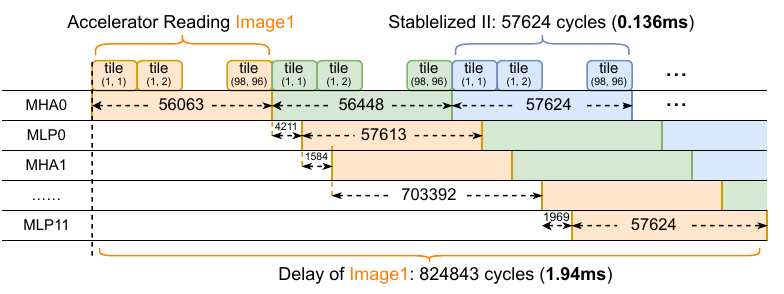}
    \caption{The timing diagram in HG-PIPE of each block.}
    \label{fig: tracing}
    \Description{The timing diagram of each layer.}
\end{figure}

\subsection{Timing Diagram of the Pipeline}
We simulated our design to generate a timing diagram of blocks, shown in Figure \ref{fig: tracing}.
%, confirming that our design achieves fully pipelined and overlapped execution.
The accelerator sequentially loads input tensor tiles. As Image1's loading completes, Image2's begins, indicating overlapped execution. The MHA block exhibits coarse-grained buffering, causing a slight delay in outputting the first tile. For the third image, the stable II measured was 57,624 cycles as expected, validating the hybrid-grained pipeline's effectiveness. The trace also reveals that the total processing time for Image1 is 824,843 cycles or 1.94ms. Using the stable II, the computed latency is 0.136 ms, equating to an ideal frame rate of 7353 images/s.

\begin{comment}
    The revised version of the table.
    The points:
    1. delete some competitors. Tomato, for it is CNN. VAQF, for it is deit-base. Via, for it is Swin-tiny.
    2. Add some competitors. SSR, for it is very similar, which also applies to the VCK190 FPGA.
    3. Add V100 result from Deit-tiny paper.
    4. Add network metrics, including parameters, and operations.
\end{comment}

\newcommand{\mr}[1]{\multirow{2}{*}{#1}}
\begin{table*}[!tb]
\centering
\footnotesize
\begin{threeparttable}
\renewcommand{\arraystretch}{1.0}
\begin{tabular}{m{1.3cm}|  m{1.3cm}  |  m{1.1cm} *{2}{m{1.5cm}} m{1.7cm} |   *{4}{m{1.2cm}}}
\hline
\hline
                                &\textbf{Deit GPU \newline Baseline} ~\cite{deit}
                                &{TCAS-I\newline 2023}   ~\cite{group_vector_sa}
                                &\textbf{AutoViTAcc}    \newline {FPL    2022}   ~\cite{auto-vit-acc}
                                &\textbf{HeatViT}       \newline {HPCA   2023}   ~\cite{heat-vit}
                                &\textbf{SSR}           \newline {FPGA   2024}   ~\cite{ssr}
                                &\multicolumn{3}{m{1.8cm}}{\textbf{HG-PIPE}  \newline     This work 2024} \\
\hline                
\textbf{Paradigm}               &GPU                    &GeMM               &GeMM                   &GeMM               &Coarse-Grained\newline Pipeline   &\multicolumn{4}{c}{\textbf{Hybrid-Grained Pipeline}}              \\
\hline
\textbf{FPGA/GPU}               &V100 GPU               &ZCU102             &ZCU102                 &ZCU102             &VCK190             &ZCU102\tnote{3}    &VCK190             &VCK190                &VCK190                 \\
\textbf{Frequency}              &1455MHz                &300MHz             &150MHz                 &150MHz             &PL:250MHz, AIE:1GHz&375MHz             &425MHz             &425MHz                &350MHz                 \\
\hline                                                                                                                                                                                                   
\textbf{Network}                &Deit-tiny              &ViT-tiny           &Deit-small             &Deit-tiny          &Deit-tiny          &Deit-tiny          &Deit-tiny          &Deit-tiny             &Deit-small             \\
\textbf{OPs/inf}                &2.5G                   &2.5G               &9.2G                   &2.5G               &2.5G               &2.5G               &2.5G               &2.5G                  &9.2G                   \\
\textbf{Params}                 &5.5M                   &5.5M               &22M                    &5.5M               &5.5M               &5.5M               &5.5M               &5.5M                  &22M                    \\
\textbf{Precision}              &fp32                   &A8W8               &A4W4+A4W3              &A8W8               &A8W8               &{A4W4}             &{A4W4}             &{A3W3}                &{A3W3}                 \\
\textbf{Accuracy}               &74.5\%                 &73.00\%            &77.94\%                &72.20\%            &-                  &74.37\%            &74.37\%            &71.05\%               &-\tnote{6}             \\
\hline                                                                                                                                                                                                   
\rowcolor{gray!20}                                                                                                                                                                                       
\textbf{FPS}                    &2529                   &245                &155.8                  &183.4              &4545               &1579               &3629               &\textbf{7118}         &1490                   \\
\rowcolor{gray!20}                                                                                                                                                                                       
\textbf{GOPs}                   &6322.5                 &762.7              &1418.4                 &366.8              &11362.5            &3947.5             &9072.5             &\textbf{17795}        &13708                  \\
\hline                                                                                                                                                                                                   
\textbf{LUTs}\tnote{1}          &-                      &114k               &193k                   &137.6k             &619k               &212.7k             &514k               &669k                  &869k                   \\
\textbf{DSPs}                   &-                      &1268               &1549                   &1968               &14405\tnote{5}     &78                 &156                &312                   &312                    \\
\textbf{BRAMs}\tnote{2}         &-                      &648                &-                      &355.5              &1456               &324.5              &1284               &1006.5 \tnote{4}      &2748                   \\
\textbf{Power}                  &-                      &29.6W              &10.34W                 &9.45W              &46W                &21.9W              &43.4W              &46.7W                 &48.1W                  \\
\rowcolor{gray!20}                                                                                                                                                                                       
\hline                                                                                                                                                                                                   
\textbf{GOPs/kLUT}              &-                      &6.69               &7.35                   &2.66               &18.35              &18.55              &17.65              &\textbf{26.60}        &{15.77}         \\
\rowcolor{gray!20}                                                                                                                                                                                       
\rowcolor{gray!20}                                                                                                                                                                                       
% normalize to DSP
\textbf{GOPs/DSP}\tnote{7}      &-                      &0.157              &0.187                  &0.058              &0.336              &0.587              &0.559              &\textbf{0.839}         &0.499                \\
\rowcolor{gray!20}                                                                                                                                                                                       
\textbf{GOPs/W}                 &-                      &25.76              &137.17                 &38.80              &246.15             &180.25             &209.04             &\textbf{381.0}        &{284.9}         \\
\hline
\hline
\end{tabular}
% \scriptsize

\begin{tablenotes}
%\tiny
%\footnotesize
\scriptsize
\item[1,2] \textbf{LUTs}: LUT-6 for AMD FPGAs. \textbf{BRAMs}: BRAM-36k for AMD FPGAs.
\item[3] The ZCU102 board is not capable of freezing all the layers of Deit-tiny, therefore we divide the network into 4 parts.
\item[4] We allocated 718.5 BRAMs and 36 URAMs to this setup. For the sake of comparison, we normalized the utilization of URAMs to BRAMs, considering 1 URAM equivalent to 8 BRAMs.
\item[5] SSR\cite{ssr} utilized 394 AI Engine (AIE) to perform the matrix multiplication, we normalized the utilization of AIE to DSPs, considering 1 AIE equivalent to 32 DSPs.
\item[6] Currently we do not have Quantization Aware Training (QAT) weight results for Deit-small or bigger models.
\item[7] \textbf{Normalized GOPs/DSP}: Since different works adopt different LUT/DSP preferences, we normalize them to be DSP-only for a fair comparison, considering 1 DSP = 32 LUTs. This is a very conservative estimation since 1 DSP is typically more capable than 32 LUTs,
and likely underestimates our true efficiency since we predominantly use LUTs. 
Nonetheless, our work still demonstrates superior efficiency compared to other works.

\end{tablenotes}
\end{threeparttable}
\caption{Comparison with prior art works.} % FPGA works
\label{tab: comparison_fpga}
\end{table*}

\subsection{Comparison with Related Works}

\ml{This is an important session. Need to make the comparison more explicit. On ZCU102, compared to XXX, \method~achieves XXX. On VCK190, compared to XXX, \method~achieves XXX.}

\sout{
We benchmarked HG-PIPE against leading works, as detailed in Table \ref{tab: comparison_fpga}. Our hybrid-grained pipeline design excelled in throughput, resource efficiency and power. With VCK190 at the highest parallelism level ($\mathtt{TP}=2$), HG-PIPE reached a throughput of 7118 images/s and 17.8 TOPs/s, nearly achieving the ideal rate of 7353 images/s (98.3\% efficiency) and outperforming the V100 GPU by 2.85$\times$ \cite{deit}. On ZCU102 with $\mathtt{TP}=1$, the LUT efficiency is 18.55 GOPs/kLUT, 2.52$\times$ of Auto-ViT-Acc on ZCU102 \cite{auto-vit-acc}, under the same 4-bit quantization and platform conditions.
}

We benchmarked HG-PIPE against leading works, as detailed in Table \ref{tab: comparison_fpga}. Our hybrid-grained pipeline design demonstrated significant improvements in throughput, resource efficiency, and power efficiency. On the ZCU102 platform, compared to AutoViTAcc, HG-PIPE achieves a LUT efficiency of 18.55 GOPs/kLUT, which is 2.52 times higher under the same 4-bit quantization and platform conditions. 
On the VCK190 platform, HG-PIPE achieved a throughput of 7118 images/s and 17.8 TOPs/s, which is 96.8\% of the ideal 7353 images/s throughput. Compared to the V100 GPU (2529 FPS), HG-PIPE outperforms it by 2.81 times. For DSP usage, HG-PIPE drastically reduced consumption to just 312 DSPs through effective LUT-based optimizations.
%, increasing DSP efficiency by an order of magnitude compared to previous works (from AutoViTAcc's 0.916 GOPs/DSP to 54.82 GOPs/DSP). 
Moreover, HG-PIPE also demonstrates great power efficiency. On the same VCK190 platform, HG-PIPE achieves 381.0 GOPs/W, outperforming prior art work SSR (246.15 GOPs/W).

\subsection{LUT Optimization Techniques}
We evaluated the impact of our LUT optimization techniques in reducing DSP usage and maintaining network accuracy in Figure \ref{fig: ablation_study_curve}. Initially, the network had 74.5\% accuracy with a prohibitive DSP cost of 14304 \cite{deit}.
Implementing advanced quantization techniques \cite{vvtq} allowed us to implement MAC units with LUTs, 
reducing DSP usage from 14304 to 3024.
Further applying Power-of-Two LUTs for non-linear functions cut DSP usage to 312, albeit with a significant accuracy loss. The crucial Inverted Exp Table technique significantly restored accuracy, with subsequent optimizations gradually recovering it while keeping DSP usage constant. We also conducted an ablation study to pinpoint the effects of optimizations detailed in Figure \ref{tab: ablation}. Also, Figure \ref{tab: resource reduction} demonstrates the table sizes and the resource reduction effects. As listed, LUT costs of non-linear functions are significantly reduced, while DSP usage is eliminated. 

% \subsection{Scalability}
% \begin{figure}[!tb]
% \centering
%     \includegraphics[width=1\linewidth]{figs/scalability.pdf}
%     % \caption{The scalability of HG-PIPE.}
%     % \caption{Scale up the design with the highly parameterizable HLS template.}
%     \caption{Scaling up HG-PIPE.}
%     \label{fig: scalability}
%     \Description{The scalability of HG-PIPE.}

% \end{figure}

% Our design is template-based and therefore highly parameterizable and scalable. As shown in Figure \ref{fig: scalability}(1), we can easily scale the design to larger networks by changing the hyperparameters, such as data types, dimensions and the corresponding parallelism. Also as shown in Figure \ref{fig: scalability}(2), this architecture can be easily extended to multiple FPGAs, for each layer performs an asynchronous pipeline without any global synchronization, and SerDes can provide enough bandwidth for the activation transmission between FPGAs. 
% % We carried a loopback test on the VCK190 board to
% To validate the cross-FPGA scalability, we split the design into two parts and carried out an external loopback test as Figure \ref{fig: scalability}(3) to emulate the cross-chip activation communication, and no performance degradation is observed.
% With the template-based design and decentralized architecture, the scalability of HG-PIPE can be guaranteed.

\subsection{Test Environment and Device View}
\begin{figure}[!tb]
\centering
    \includegraphics[width=1\linewidth]{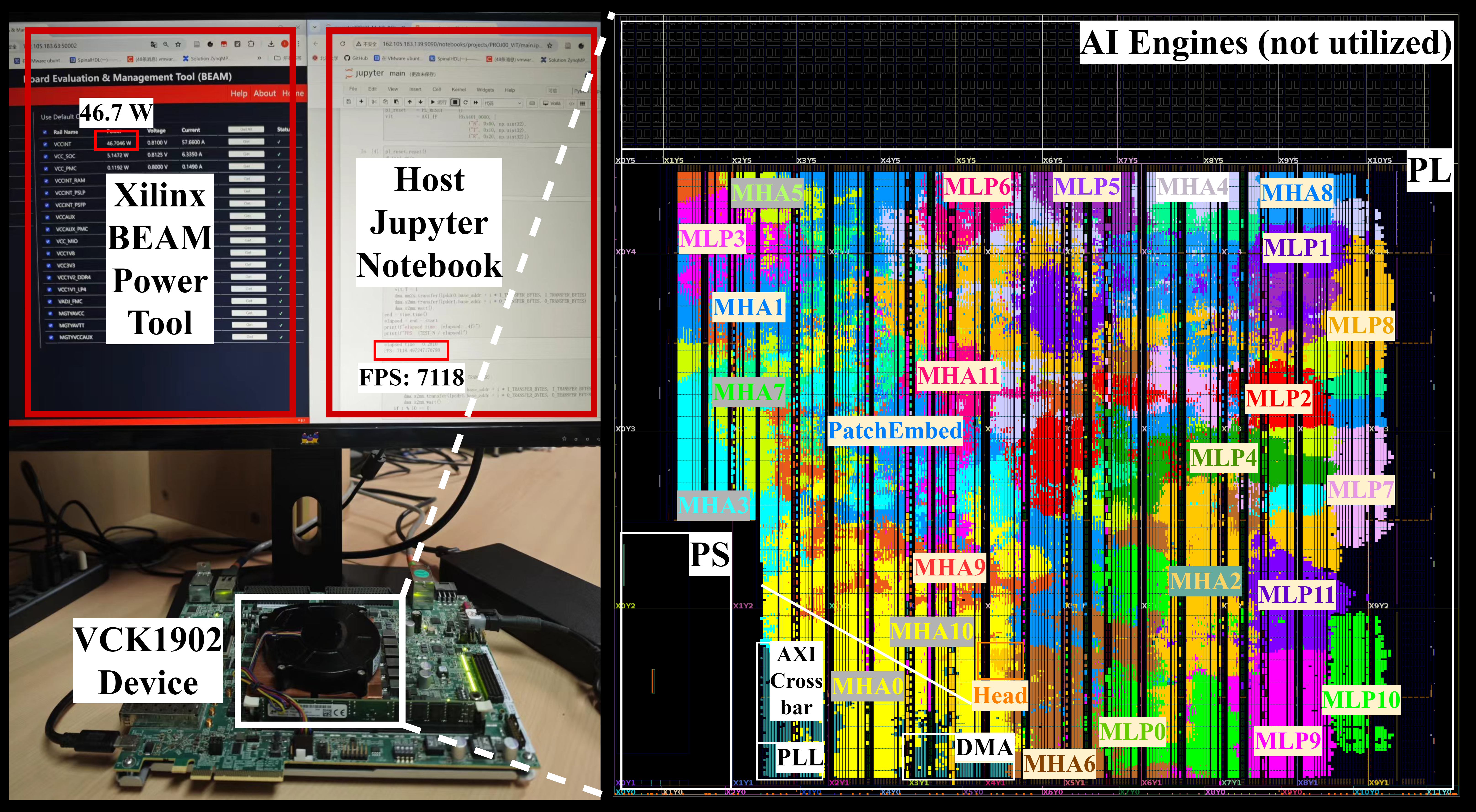}
    \caption{Test environment and device view of HG-PIPE.}
    \label{fig: device_view}
    \Description{The test environment and device view of the HG-PIPE on VCK190.}
\end{figure}
Our test environment is demonstrated in Figure \ref{fig: device_view}. The VCK190 board is connected to the host PC via Ethernet. The screen of the host PC displays the BEAM and Jupyter Notebook GUI, highlighting real-time power consumption and frame rate. On the right side, the implementation device view is elaborated with each module annotated. The device view contains 26 neural network blocks, including PatchEmbed, Classification Head, 12 MHA blocks and 12 MLP blocks. The placement is automatically generated by Vivado, thus appearing in a random order. Other modules such as PS, PLL, DMA and AXI Crossbar are also annotated. Notably, no AI Engine is utilized and all computation is performed on the PL side.

\section{Conclusion}
In this paper, we introduce \method, a hybrid-grained pipelined ViT accelerator optimized for FPGAs. By combining the advantages of fine-grained and coarse-grained designs, our approach achieves low latency and high resource efficiency at the same time, reducing the on-chip activation buffering cost by 83.3\%. 
We have implemented optimizations including PoT table index computation and LUT-based ReQuant, which collectively reduce DSP usage by 89.6\% without sacrificing accuracy. 
% Tested on ZCU102 and VCK190 FPGAs, \method outperforms existing accelerators, achieving 2.72$\times$ higher throughput and 2.46$\times$ better resource efficiency on ZCU102. 
On the VCK190, it delivers real-time ViT processing at 7118 FPS, equivalent to 17.8 TOP/s. 
% With the scalability guaranteed by the template-based design, HG-PIPE can be easily extended to support larger ViT models and more FPGA platforms.


%% \begin{acks}
%% To Robert, for the bagels and for explaining CMYK and color spaces.
%% \end{acks}

% \newpage
\printbibliography

% %%
% %% If your work has an appendix, this is the place to put it.
% \appendix









\end{document}